# Computer-aided whole-cell design: taking a holistic approach by integrating synthetic with systems biology

Lucia Marucci[1,2,3,+], Matteo Barberis[4,5,6,*,+], Jonathan Karr[7,*,+], Oliver Ray[8,*,+], Paul R. Race[9,3,*,+], Miguel de Souza Andrade[10,11], Claire Grierson[12,3,+], Stefan Andreas Hoffmann[13], Sophie Landon[1,3], Elibio Rech[10,+], Joshua Rees-Garbutt[12,3], Richard Seabrook[14,+], William Shaw[15], Christopher Woods[16,3,+].

[1]Department of Engineering Mathematics, University of Bristol, BS81UB, Bristol, UK
[2]School of Cellular and Molecular Medicine, University of Bristol, BS81TD, Bristol, UK
[3]Bristol Centre for Synthetic Biology (BrisSynBio), University of Bristol, BS81TQ, Bristol, UK
[4]Systems Biology, School of Biosciences and Medicine, Faculty of Health and Medical Sciences, University of Surrey, GU2 7XH Guildford, Surrey, UK
[5]Centre for Mathematical and Computational Biology, CMCB, University of Surrey, GU2 7XH Guildford, Surrey, UK
[6]Synthetic Systems Biology and Nuclear Organization, Swammerdam Institute for Life Sciences, University of Amsterdam, 1098 XH Amsterdam, The Netherlands
[7] Icahn Institute for Data Science and Genomic Technology, Department of Genetics and Genomic Sciences, Icahn School of Medicine at Mount Sinai, 1255 5th Avenue, Suite C2, New York 10029, US
[8]Department of Computer Science, University of Bristol, BS81UB, Bristol, UK
[9]School of Biochemistry, University of Bristol, BS8 1TD, Bristol, UK
[10]Brazilian Agriculture Research Corporation/National Institute of Science and Technology - Synthetic Biology, Brasília, DF, 70770-917, Brazil
[11]Laboratório de Biologia Teórica e Computacional (LBTC), Universidade de Brasília DF, Brazil
[12]School of Biological Sciences, University of Bristol, BS8 1TQ, Bristol, UK
[13]Manchester Institute of Biotechnology, University of Manchester, 131 Princess Street, M1 7DN, Manchester, UK
[14]Elizabeth Blackwell Institute for Health Research (EBI), University of Bristol, BS8 1UH, Bristol, UK
[15]Department of Bioengineering, Imperial College London, London SW7 2AZ, UK
[16]School of Chemistry, University of Bristol, BS8 1TS, Bristol, UK

[*] These authors contributed equally to this work
[+] Correspondence should be addressed to L.M. (lucia.marucci@bristol.ac.uk), C.G. (claire.grierson@bristol.ac.uk), minimal genome design; M.B. (m.barberis@surrey.ac.uk), network design and multiscale model derivation; O.R. (csxor@bristol.ac.uk), symbolic modelling; J.K. (karr@mssm.edu), whole-cell model development; P.R.R., E.R., R.S. (Paul.Race@bristol.ac.uk, elibio.rech@embrapa.br, richard.seabrook@bristol.ac.uk), bioindustrial applications; C.W.(Christopher.Woods@bristol.ac.uk), software engineering.

**Abstract**
Computer-aided design for synthetic biology promises to accelerate the rational and robust engineering of biological systems; it requires both detailed and quantitative mathematical and experimental models of the processes to (re)design, and software and tools for genetic engineering and DNA assembly. Ultimately, the increased precision in the design phase will have a dramatic impact on the production of designer cells and organisms with bespoke functions and increased modularity. Computer-aided design strategies require quantitative representations of cells, able to capture multiscale processes and link genotypes to phenotypes. Here, we present a perspective on how whole-cell, multiscale models could transform design-build-test-learn cycles in synthetic biology. We show how these models could significantly aid in the design and learn phases while reducing experimental testing by presenting case studies spanning from genome minimization to cell-free systems, and we discuss several challenges for the realization of our vision. The possibility to describe and build *in silico* whole-cells offers an opportunity to develop increasingly automated, precise and accessible computer-aided design tools and strategies throughout novel interdisciplinary collaborations.

# 1. Introduction

Whole-cell models (WCMs) are state-of-the-art Systems Biology formalisms: they aim at representing and integrating all cellular functions within a unique computational framework, ultimately enabling a holistic and quantitative understanding of cell biology (1, 2). Quantitative and high-throughput *in silico* experiments generated from WCMs promise to significantly shorten the distance between hypothesis/design formulation and testing (3).

While simplified models for specific cellular functions were first developed over 30 years ago (e.g. formalisms for gene expression regulation (4), signaling (5) and metabolic (6) pathways, cell growth (7) and the cell cycle (8-10)), the first WCM, namely the E-Cell model, was only derived in the 1990s for *Mycoplasma genitalium*, the smallest living organisms with a 580 kb genome, following publication of the organism genome sequence (11). The so-called virtual self-surviving cell (SSC) partially stochastic model includes only a subset of protein-coding genes (105) and enables dynamic simulations which encompass various subcellular processes, including enzymatic reactions, complex formation and substance translocation. In parallel, the first genome-scale metabolic models (GSMMs) were developed by Palsson's group (12) using flux balance analysis (FBA) in the 1990s.

More recently, hundreds of GSMMs have been reconstructed for different organisms, with an increasing number of represented genes (13-15). GSMMs have been complemented with a mathematical description of other processes, such as transcription, translation and signaling (16, 17). Less than a decade ago the first hybrid dynamic model, representing all genes and molecular functions known for an organism, was proposed by Covert's group (18). In this pioneering work, Karr and colleagues integrated 28 sub-models within a unique MATLAB WCM to represent one cell cycle of *M. genitalium*; each sub-model is represented with a distinct formalism, including ordinary differential equations (ODEs), flux balance analysis (FBA), stochastic simulations and Boolean rules.

Substantial research and effort are still needed to improve WCMs' descriptive power and to increase the complexity of organisms they can represent. A complete WCM should ideally integrate multiscale interactions at the cellular level (18, 19) while accounting for the overall cellular structure (20), the dynamic structure of molecular interactions (21-23) and the spatial compartment of the subcellular components (24-26). Ensuring an accurate representation of all the mentioned processes is highly challenging, especially when aiming to model increasingly complex organisms (27-29). We refer the reader to recent efforts which provide an overview of the state-of-the-art in the development of WCMs (30, 31).

Here, we focus on the enormous potential we believe WCMs have for design-build-test cycles integrating synthetic with systems biology (Figure 1). While the discussed applications are diverse, they share a high degree of complexity which would imply extensive trial and error experimental cycles in the absence of robust computational design algorithms based on predictive models. We conclude by considering relevant challenges that interdisciplinary communities should address to fully realize our vision, discussing future directions for the integration of WCM development through synthetic and systems biology applications.

# 2. Whole-cell design strategies in synthetic biology
## 2.1 Model granularity of gene network (re)design

Mathematical models can be instrumental to (re)design network circuits that recapitulate definite biological functions. Knowledge of regulatory mechanisms in biological pathways has been gained by considering living systems as a composition of functional modules, which are investigated through minimal computer models. Examples include controllable oscillators (32-35), circadian clocks (36, 37), signaling networks (38), the metabolism (39, 40) and transcriptional regulation (41). The granularity of biochemical details uncovered about a definite regulatory network may differ between minimal and detailed computer models.

However, one may expect that if the core design of a minimal and a detailed model is similar, their general properties will match.

Understanding of a living organism at a system's level may be reached through decomposing a system into functional modules, or modular circuits that perform a definite task (42-44). The capability to sustain viability through autonomously generated offspring is essential for organismal functions and, therefore, is a feature WCMs shall account through modeling cell division, which is intimately integrated with various layers of cellular regulation (metabolism, signaling, gene regulation, transcription, etc.). A number of minimal models have been developed for the eukaryotic cell cycle by Barberis's, Tyson's and Novák's groups (45-50).

Currently, the majority of multiscale models (not WCMs) lack components able to bridge cellular networks/function (cell cycle, metabolism, signaling, gene regulation, etc.). Identification of hubs, i.e. elements with high connectivity in the cellular environment that integrate cellular networks, is a critical feature of WCMs. Transcription factors have recently been identified as hubs that integrate multiscale networks, potentially connecting the cell cycle to metabolism (51), and can be among the parts of a system that influence its state as a whole. Multiscale frameworks coupling a variable granularity of these networks are being developed, by identifying the relevant regulations occurring among common network nodes and through the use of different mathematical formalisms (52). These and other strategies are developed to progressively integrate networks of other cellular functional modules (53). Together with the identification of network designs underlying cell's autonomous oscillations, these strategies can rationalize the proper timing of offspring generation accounted by WCMs.

Designing synthetic gene networks by integrating them within WCM formalisms (as in (33)) could be critical to investigate how gene expression correlates with codon usage, explore possible cell burden effects (54), and predict modularity of synthetic gene networks and tools to modulate gene expression across different chassis (55-57).

2.2 Design and engineering of reduced genomes
Minimal genomes can be defined as reduced genomes containing only the genetic material which is essential for a cell to reproduce (58). Studying and engineering minimal genomes can be instrumental both to understand the most essential tasks a cell must perform to sustain life, and to obtain optimal chassis for synthetic biology applications, with reduced cell burden and superior robustness (59-63). Genome-scale computational models of cells can be instrumental to fully understand the dynamic and context-dependent nature of gene essentiality (64), and to rationally design minimized genomes *in silico*. Computer-aided minimal genome engineering could significantly reduce the time and cost of approaches used so far to produce reduced genomes, which can otherwise be generated by extensive experimental iterations (60, 65-70).

To the best of our knowledge, two top-down genome reduction approaches have been proposed so far based on computational genome-scale models. The MinGenome algorithm applies a mixed-integer linear programming (MILP) algorithm to a GSMM of *E. coli*, using information pertaining to essential genes and synthetic lethal pairs within the optimization (71). In contrast, Minesweeper and GAMA are top-down genome minimization algorithms based on the *M. genitalium* WCM. They exploit a divide-and-conquer approach and a genetic algorithm, respectively, to iteratively simulate reduced genomes (72); their *in silico* predictions have not been tested in the laboratory yet.

Applying GSMM-based genome reduction algorithms such as MinGenome or analogous metaheuristic equivalents which could be easily adapted for this aim (e.g. (73-75) is, at the current stage, more broadly applicable across organisms given the large availability of these formalisms. Still, as more WCMs become available, we expect WCM-based genome reduction algorithms to provide superior predictions of cellular processes and genetic interactions, thanks to their richness of multiscale cellular process representation.

2.3 Design and prototyping of cell-free systems
Cell-free transcription/translation systems, based on crude cellular extracts, are a valuable platform to address fundamental biological questions in a controllable and reproducible way. In recent years, the decrease of costs associated with this technology and significant improvements in synthesis yield capabilities (76) have made cell-free systems increasingly popular in synthetic biology for the prototyping and testing of engineered biological parts (13-15, 70) and networks (77-79). As the possible applications of cell-free systems grow (see (80) for a recent review where a spectrum of cell-free probing of cellular functions and of applications are discussed in detail), mathematical models are developed to quantitatively formalize how biological processes perform within cell-free platforms (81).

So far, deterministic models (ordinary differential equations-, or constraint-based) have been proposed to describe specific processes within cell-free platforms such as transcription and translation (78, 82, 83), resource competition (84-87) and metabolism (88). The integration of mathematical formalisms across scales for cell-free platforms, building towards WCMs, could be highly beneficial to both facilitate *de novo* design of circuits, and to quantitatively compare *in vitro* cell-free products with their *in vivo* counterparts.

2.4 Whole-cell biosensor design and testing
Biosensors are analytical devices which can convert a biochemical reaction into a measurable signal. The recognition unit in a biosensor can be composed of whole cells, nucleic acids, enzymes, proteins, antibodies or combinations thereof. Synthetic biology has significantly accelerated biosensor development; new generation whole-cell biosensors (i.e. sensors implemented throughout living cells) have been engineered, for example, to detect arsenic (89) and other analytes (including pollutants and antibiotics (90)), for microbial detection in industrial settings (91) and for *in vivo* diagnostic applications (e.g. detection of environmental signals in the gut (92) and diagnosis of liver metastases (93); see (94) for an overview of translation of promising technologies into diagnostic tools).

The application of WCMs to the design, prototyping and testing of whole-cell biosensors could suggest rational approaches to tune their sensitivity, stability and dynamic range while facilitating the choice of the ideal chassis and, if needed, guide its re-engineering to optimize biosensor performance (95). If WCMs become available for different chassis and entire organisms, they could also support the design of optimized targeted delivery of genetically encoded biosensors.

2.5 Industrial implications of whole-cell models
Although the intellectual merit of pursuing a computer-aided whole-cell design approach is unquestioned, it is clear that the success of this endeavor will ultimately be judged by its impact on industry. The increasing drive towards 'green' chemistry approaches, allied to increases in gene synthesis speed and capability and associated cost reductions, are making biosynthesis an increasingly appealing route for the manufacture of high-value chemicals (96). This includes a plethora of opportunities across the pharmaceutical, agrochemical, commodity chemical, and materials sectors, amongst others.

A major challenge, however, remains the development of robust, scalable microbial chassis, whose metabolic processes can be predictably tuned for a desired outcome (97). Currently, chassis choice is largely restricted to a subset of genetically tractable microorganisms, whose physiology and performance during fermentation are well understood, and for whom effective molecular genetic tools required for their manipulation exist. Chassis optimization to date has relied exclusively on incremental, stepwise improvements in desired host strain characteristics, including growth rate, feedstock utilization and product yield (98). For these reasons, the process of chassis optimization remains prohibitively slow and expensive, accounting in part for the paucity of high-value small molecules that are currently manufactured using synthetic biology processes. Targeted manipulations often lead to unanticipated off-target effects, linked to the co-dependency of metabolic processes, which generally function in concert within interdependent cellular networks (99): perturbations may compromise rather than enhance desirable characteristics, leading to undesired outcomes. Clearly, robust, predictable WCMs represent an attractive solution to the problem of chassis optimization, affording a catch-all tool that can be used to unpick dependencies and ensure that performance criteria can be met. Besides, the complexities associated with population heterogeneity during chassis fermentation must be resolved (100). For fermentation-based industrial processes to be tractable, product yields must be sufficiently high to make biosynthesis financially viable. The emergence of 'cheaters' or slow-growers within microbial populations must be addressed. This will undoubtably be best achieved through tunable regulatory processes that operate throughout populations. The introduction of such characteristics is a major challenge to conventional chassis design approaches. However, this could be much more easily implemented and tested by employing WCM driven approaches.

Critical to the success of a computer-aided whole-cell design approach is the quality of the employed model (101). Microbial systems with small genomes represent a compelling entry point for study, with model development possibly being facilitated by ongoing studies focused on establishing the core constituents of a functional genome. These studies are being predominantly driven by genome minimization experiments, which in turn can be used to further refine model performance. Importantly, fundamental gaps remain in our understanding of microbial metabolic processes, and this will unquestionably hinder progress (102). However, the capacity of WCMs to predict previously unidentified metabolic dependencies should be viewed as an acid test of model validity. Indeed, GSMMs often fail due to their inability to account for interdependencies, a feature which has led to skepticism within industrial circles, questioning the value of such models. Whole-cell approaches offer a mechanism to circumvent this issue and could increase reliability and promote confidence in the adoption of WCMs. This is of particular significance when developing chassis for 'non-natural' products whose chemistries sit outside those of metabolites found in nature (98). Expanding the metabolic capacity of chassis organisms to deliver such novel products risks introducing additional complexities, including excessive depletion of core metabolite pools or the generation of toxic products or intermediates. Design approaches driven by WCMs are uniquely placed to identify such issues and provide a route to their circumvention.

The capacity to design-in explicit control over cellular behavior is also critical for industrial adoption of model-derived chassis. It can be argued that the ability to regulate cellular processes is as important as defining the processes themselves. Tunable regulatory systems must afford a degree of both intrinsic and extrinsic control. Synthetic biology-based approaches for the construction of genetic circuitry are now placing us on a path to broad-reaching cellular regulation, though issues still exist. These systems are often insufficiently orthogonal, with bespoke designs required for different chassis due to variations in core metabolic process (39). Again, whole-cell design approaches offer a solution to this issue, as such systems can be predefined and tested for functionality *in silico* prior to undertaking costly lab experimentation.

## 3. What's next? Going beyond the prototype

In recent years, advances in genomic measurement technologies for data generation, the establishment of data repositories, and the development of WCM simulation platforms have significantly facilitated the derivation of WCMs (see (30) for a review). Nevertheless, the implementation of WCM-based design-build-test cycles for genome-scale engineering requires further challenges to be addressed (103).

If a model has to be used for the design and prototyping of an engineered living system, the model needs to be reliable. Even for a simple organism, the number of kinetic parameters surges as the complexity and the level of detail of a mathematical model increase; constraining parameters thus becomes harder and requires extensive experimental data. To set the 1,462 quantitative parameters of the *M. genitalium* WCM, values from other species were incorporated due to a lack of organism-specific data (104); a combination of parameter values reported from previous experiments and numerical optimization on a reduced model was performed. While, ideally, we would like to measure all kinetic parameters directly from experiments, we still lack techniques to measure each state in individual cells over time, and across all possible environmental conditions. Mathematical models can be used to produce predictions of missing data, however they often abstract physical processes using simplifying assumptions which might hold in specific conditions only (105). A combination of direct experimental estimation and parameter inference might be needed for genome-scale formalisms and WCMs, for example using scalable inference techniques to parameterize sub-models.

Sensitivity analysis, usually performed by perturbing parameters to understand how uncertainties affect the model outputs (106), can become extremely computationally expensive when applied to genome-scale models. Alternatively, statistical approaches such as those based on Bayesian methods (107) or on the Fisher information matrix (108) could be carefully carried out at least at the sub-model level, and possibly scaled up to WCMs. The Reverse Engineering Assessments and Methods (DREAM8) parameter estimation challenge (109) was organized to develop new parameter estimation techniques specific for WCMs. It suggested possible interesting avenues for WCM parameterization (i.e. model reduction and a combination of differential evolution and random forests), and highlighted that the availability of comprehensive data-set is critical to ensure the model is practically identifiable (110).

It is also important to consider the structural uncertainties in the model, which depend on model assumptions. While, for certain sets of models (e.g. small ordinary differential equation systems for signaling pathways), likelihood- and Bayesian-based approaches have been proposed for model selection (111) (112) and Semidefinite Programming for model invalidation (113), no suitable techniques for WCM selection have been proposed to date.

We foresee that automation will play a fundamental role in the derivation of WCMs for eukaryotic organisms and in their application to design complex processes. Ideally, we would like to introduce automation at different stages, such as data extraction from the literature, model derivation, and model/data integration both within the model fitting and validation steps, and when comparing *in vitro* design prediction with *in vivo* tests (103). This, in turn, will require the adoption of standards for both data and model repositories.

Researchers have started to collect data needed for WCM development into public repositories (e.g. (114-118)); still, data needed to derive and fit WCMs are dispersed across many repositories and publications and often not annotated or normalized, ultimately requiring a massive manual effort. Federated archives of repositories, such as the PDB-Dev system to deposit Integrative/Hybrid models and corresponding data (119), might be well placed to archive and disseminate both data and models, while enabling different researchers to attempt alternative modelling/parameterization approaches. Covert's group developed the WholeCellKB database (120) to organize the quantitative measurements (over 1,400) from which the *M. genitalium* WCM was derived; it would be ideal to enable automatic access and querying in such databases.

To enhance WCM reproducibility, new standards and simulations software are also needed (121). Researchers should invest efforts to use standard formats such as the Systems

Biology Markup Language (SBML) (122) and the Systems Biology Graphical Notation (SBGN) (123). At present, various aspects of the *M. genitalium* WCM cannot be represented by SBML, mainly due to the multi-algorithmic nature of the model itself (124). Further development of standard modelling formats is needed to enable reproducible WCM simulations, e.g. by including in the SMBL Hierarchical Model Composition package ontologies which could represent the algorithm needed for specific sub-models (125). In the context of synthetic biology applications, we believe it would be appropriate and beneficial to report and deposit data related to various iterations of WCM-generated *in silico* predictions, *in vivo* testing and possible model/design refinement; this would establish the predictive power of WCMs and illuminate steps to make design-build-test-learn cycles more effective.

To facilitate the adoption of WCMs for synthetic biology applications, high-performance parallelized computer clusters are required to coordinate and run the models with lengthy runtimes and large corresponding databases across computer clusters, to parameterize and validate the models, and then to integrate them in design cycles in combination with optimization algorithms (104) (126).

The implementation of standardized tools to share data and simulate WCMs would, in turn, facilitate model validation. This should involve the definition of proper metrics and formal model verification techniques such as those developed for SBML-encoded models (127).

## 4. (Re)thinking system approaches: a collaborative effort

In addressing the aforementioned challenges, we believe there is a tremendous opportunity to rethink approaches used so far to generate genome-scale models, including WCMs, and to integrate with broader communities including software engineers, computer scientists, structural biologists, bioinformaticians, and systems and synthetic biologists.

We do anticipate that, as diverse communities synergize on WCM-related research, different kinds of formalisms might be integrated within genome-scale models. Symbolic reasoning (e.g. satisfiability solving, model checking, theorem proving, formal methods, logic programming and Boolean networks) provide a range of expressive and intuitive logical frameworks that could potentially complement and help glue together sub-models at different scales. Such methods are routinely applied to complex systems in the electronics and software industries, and have been tentatively applied to biological systems for nearly a decade (128). Recent work showed the feasibility of applying logic programming methods to signaling pathways (129), metabolic networks (130) and automating a mechanistic philosophy of scientific discovery in simulated organisms (131); it should be feasible to integrate such sub-models within a WCM framework.

We believe there is scope to further increase the descriptive and predictive ability of WCMs across spatial and temporal scales by integrating the structural biology and the molecular modelling communities to carefully consider not only the biochemical, but also the physical, molecular and structural components of cells. The development of the so-called "physical" WCMs (see (31) and (132) for comprehensive reviews) is an emerging field, with the first models describing minimal cellular environments in full atomistic detail (21, 133). With the final aim to integrate cellular and physical WCMs within a multiscale framework (134), we need approaches which can cope with the limitations of atomistic models of biomolecules (mainly in terms of computational resources), possibly exploiting coarse-grained (135, 136) or continuum (137) approaches.

By collaborating with software engineers, we need to develop tools which can enable, and possibly automate, the integration of different data types across scales, model derivation, fitting and validation, and visualization and interpretation of results (27). Rule-based models might become the new standard to represent each molecular species with the required level of granularity and multi-algorithmic sub-models (e.g. flux balance analysis and stochastic dynamical models). Frameworks where intuitive logic is coupled to rule-based models have started to be developed recently (52).

As we produce ever-increasing amounts of experimental data and increasingly sophisticated computational tools to realize detailed and complex representations of actual cells, approaches instead focusing on deliberately abstract and parsimonious simulations of artificial cellular systems provide a valuable change of perspective. Such "toy models" might be a valuable tool to test different algorithms for model derivation and fitting, while offering an opportunity to engage with broader research communities and with the public (138).

Finally, we believe there is tremendous potential for applying machine learning techniques to both WCM derivation and their applications in synthetic biology. Two recent works (139, 140) showed that deep neural networks are well placed to reconstruct the architecture of living systems (namely, the hierarchical organization of nuclear transcriptional factors in the nucleus (139) and of a basic eukaryotic cell (140)) and predict cell states and phenotypes. In both cases, the configuration of network layers and thus the biological structure were formulated using extensive prior knowledge, ultimately enabling fully "visible" systems, where all the internal biological states can be interrogated mechanistically (141). Machine learning could be beneficial to systematically process large *in vivo* and *in silico* whole-cell data-sets, for example by applying Bayesian inference, to integrate data from diverse sources and supplement sparse data (142), and to help to automatically classify WCM simulations and link phenotypes to genotypes (143). Optimal experimental design techniques might also offer a valuable methodology to select the best experimental datasets for both model identification and validation (144).

## 5. Discussion

We have shown that WCMs are likely to be instrumental to inform design-build-test cycles across synthetic biology applications. WCMs can accelerate the realization of "designer" cells and organisms tailored to specific functions, reducing experimental iterations and increasing the predictive power of computational formalisms used so far.

In the (re)design of cellular network functionalities, it is therefore important to quantitatively analyze and predict, through dedicated modelling strategies, the dynamics of interactions between various layers of cellular regulation. Thus, WCMs shall take into account how different cellular layers are integrated, and how regulatory feedback among these layers occurs in time. These challenges are tackled through integrative computational and experimental collaborative efforts aimed, respectively, towards: (i) engineering *in vivo* network designs which, through predictive systems biology, may be able to autonomously oscillate, sustaining generation of offspring, and (ii) extraction, visualization and functional exploration of regulatory interactions among cellular layers through novel multiscale modeling frameworks.

As synthetic biology moves toward the (re)engineering of entire genomes and multicellular systems, interdisciplinary communities need to collaborate for the development of tools that are required to improve the predictive power of WCMs. Although challenges remain, it is clear that the adoption of model-based methods has the potential to transform both basic research and the current bioproduction development process, leading to marked improvements in host performance and product yield on an industrial scale.

Ultimately, as the development of human genome-scale kinetic models becomes more feasible (27, 145), whole-cell formalisms might become an indispensable tool to study human variation, and design bespoke treatments and synthetic cellular screening systems.

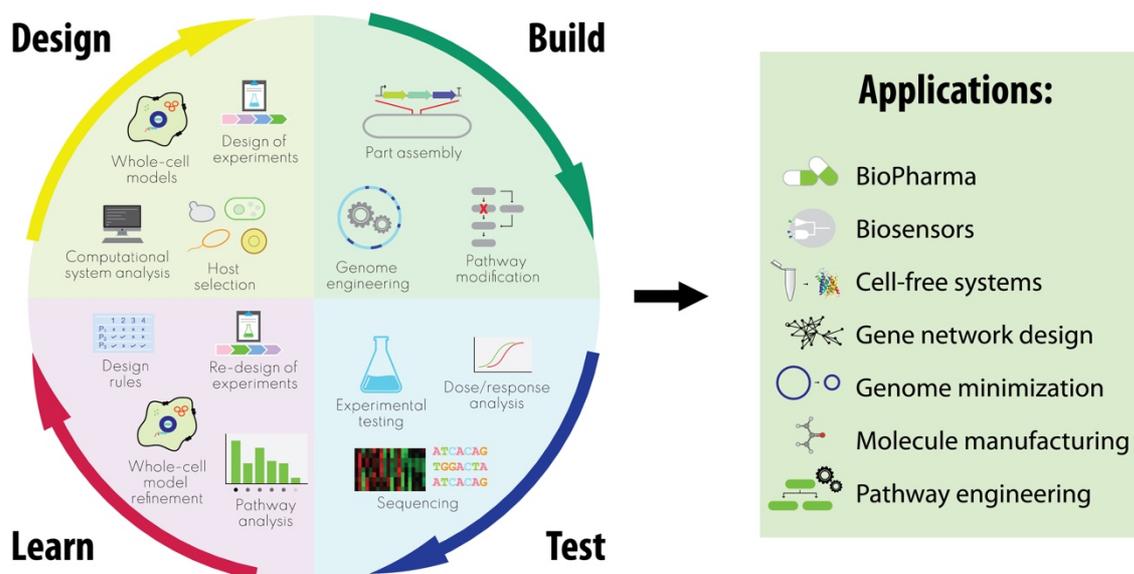

**Figure 1** Integrated design-build-test-learn cycles in synthetic biology encompassing whole-cell model-guided approaches, and relative applications.

**Author contributions**
L.M., M.B., J.K., O.R. and P.R.R. wrote the manuscript; M.d.S.A. prepared the figure; all other authors participated to discussion within the workshop, helped with editing and/or provided feedback.

**Acknowledgements**
This work captures discussions between participants at the "Computer-Aided Whole-Cell Design and Engineering" Workshop held on the 02-03 July 2019 at the University of Bristol, UK, and funded by the Engineering and Physical Sciences Research Council (EPSRC) within the remits of the Big Ideas initiative. We sincerely thank Dr Kathleen Sedgley for her support with the workshop organization, and Dr Thomas Gorochowski for participating in discussions.

**Funding**
L.M. was funded by the Engineering and Physical Sciences Research Council (EPSRC, grants EP/R041695/1 and EP/S01876X/1) and Horizon 2020 (CosyBio, grant agreement 766840); M.B. was funded by the Systems Biology Grant of the University of Surrey; J.K. was funded by the National Institutes of Health (award R35GM119771); P.R.R. was funded by EPSRC (EP/R020957/1) and the Biotechnology and Biological Sciences Research Council (BBSRC, BB/T001968/1); C.G., L.M. and P.R.R. were funded by BrisSynBio, a BBSRC/EPSRC Synthetic Biology Research Centre (BB/L01386X/); S.L. and J.R. were funded by EPSRC Future Opportunity PhD scholarships; E.R. was funded by INCT BioSyn (National Institute of Science and Technology in Synthetic Biology), CNPq (National Council for Scientific and Technological Development), CAPES (Coordination for the Improvement of Higher Education Personnel), Brazilian Ministry of Health, and FAPDF (Research Support Foundation of the Federal District), Brazil.

**Conflicts of interest**
The authors declare no competing financial interests.